\documentclass{ws-ijmpa}

\begin{document}

\markboth{R.Delbourgo and P.D.Stack}
{GR of chiral-property}
\catchline{}{}{}{}{}

\title{The Relativity of Chiral Property}

\author{Robert Delbourgo and Paul D Stack}

\address{School of Physical Sciences, University of Tasmania, 
Locked Bag 37 GPO\\ Hobart, Tasmania 7001,AUSTRALIA 7001\\
bob.delbourgo@utas.edu.au, pdstack@utas.edu.au}

\maketitle
 
\begin{abstract}
The standard model ascribes distinct properties to different chiralities of fermions.
We show how to incorporate this aspect in an extended spacetime-property framework
involving two different attributes using a generalized metric which includes gauge fields
as well as gravitation. Because the gauge fields are accompanied by coupling constants,
all such schemes, including ours, necessitate coupling unification at high energy
to ensure universality of gravtitational interactions with matter.
\end{abstract}
\maketitle

\section{\label{sec1}Chiral attributes}
The electroweak part of the standard model\cite{G,SW,W,S} unequivocally informs us 
that different fermion chiralities carry distinct properties, when gauged. The weak 
isospin ($I$) and hypercharge ($Y$) assignments of the chiral parts of
leptonic doublets are carefully designed so that the charge $Q=I_3 +Y/2$
emerges correctly; however the significant point is that right and left components
transform very differently under the associated gauge groups. The aim of
this paper is to see how these features can be incorporated with gravity
in a scheme based on {\em spacetime} coordinates augmented by Lorentz
scalar anticommuting {\em property} coordinates. 

In two earlier papers\cite{DS,SD} we investigated such a scheme but paid scant attention 
to parity as we were only concerned with the unification of gravity with 
electromagnetism  (using one property, charge)  or with Yang-Mills theory 
(two isotopic  properties). We showed how such a unification could be 
achieved, with the curvature arising from property. In particular, the Lagrangian
and energy momentum tensor of the gauge fields that transfer property
across spacetime were automatically generated. The chiral fermions were treated 
uniformly and assumed to have the same attributes in those works. It was clear, 
but more complicated, how to generalize the process for three properties, as 
needed in QCD, but we shied from carrying out the algebra as it led to a plethora of 
gauge invariant curvature coefficients in the metric that we are still uncertain how 
to constrain. With only two properties we have however the full machinery needed to 
tackle the salient properties of electroweak physics, just by focussing on the subgroup
U(1)$_L \times$ U(1)$_Y$,  confining ourselves to the charged lepton interactions. 
Before doing so we shall quickly recapitulate the way this is normally handled 
and our unification goal, if nothing else to establish our conventions; this we do in 
Section 2. We then turn to the implementation with anticommuting scalar attributes 
in Section 3, starting with the fermions which are, after all, at the root of the entire 
gauge group; then we turn to the scalar supermultiplet, which includes the uncharged 
Higgs boson\cite{EB,H,GHK,K}.

The way that gravity and the gauge bosons enter is through an extended 
spacetime-property metric (obtained from frame vectors) which incorporates 
the correct gauge group, and this is described in Section 4. Having achieved 
the goal of combining chirality with anticommuting attributes and 
curved spacetime, we offer some conclusions in Section 5 and what to expect when 
generalizing to the full SU(3)$\times$SU(2)$_L$$\times$U(1)$_Y$ of the standard
model. The most significant point of our analysis is the conclusion that any 
scheme, such as ours, where gravity and other forces are married through some
extended metric must lead to a single force strength at some large scale;
extrapolation to low scales via the renormalization group has to then result
in the known different forces, given the particle spectrum. Such are the 
overriding demands of gravitational universality.

\section{\label{sec2}Charged leptons}
Before launching into our property scheme for dealing with chiral fermions,
let us describe the endpoint attained by conventional theory,
ignoring neutrinos and charged vector bosons. We are dealing therefore with the 
U(1)$_L\times$U(1)$_Y$ subgroup of the electroweak gauge group; this does at
least include electromagnetism and the weak neutral current interactions of
the charged leptons. In preserving the same charge $Q=I_3 +Y/2$ for
the left and right pieces of the fermion field under local gauge transformations,
\begin{eqnarray}
\psi_L&\rightarrow&e^{-i(\vartheta+\vartheta')/2}\psi_L{\rm ~for~} I_3=-1/2, Y=-1\nonumber\\
\psi_R&\rightarrow& e^{-i\vartheta'}\psi_R; \qquad{\rm ~~for ~} I=0, Y=-2,
\end{eqnarray}
one encounters the covariant derivatives:
\begin{equation}
D\psi_L = [\partial +i(gW+g'B)/2]\psi_L,\quad D\psi_R=[\partial+ig'B]\psi_R,
\end{equation}
where $W, B$ are of course the associated gauge fields with couplings $g,g'$.
Hence the only way to induce a mass term in addition to the usual kinetic pieces
is to introduce a Higgs-like scalar $h$ which transforms inversely as
\begin{equation}
h \rightarrow e^{i(\vartheta'-\vartheta)/2}h,
\end{equation}
so as to ensure invariance of the Yukawa interaction,
\[\overline{\psi_L}h\psi_R +  \overline{\psi_R}h^\dag\psi_L .
\]
The gauge invariant kinetic motion of the scalar field,
\begin{equation}
Dh^\dag.Dh= [\partial+i(gW-g'B)/2]h^\dag.[\partial-i(gW-g'B)/2]h,
\end{equation}
can then provide a mass to the gauge field combination $gW-g'B$, provided
that the scalar field acquires a vacuum expectation value $\langle h\rangle
= v \neq 0$, by some mechanism.

This is where the weak mixing angle $\theta$ comes in\cite{SW,W}; let 
$\tan\theta=g'/g$ and as usual define $e=g\sin\theta=g'\cos\theta$, whereupon the
orthogonal fields $A$ and $Z$ emerge from the inverse rotation,
\begin{equation}
\left( \begin{array}{c} A\\Z \end{array}\right) =
\left(\begin{array}{cc}\cos\theta&-\sin\theta\\ \sin\theta &\cos\theta \end{array}\right)
\left(\begin{array}{c} B\\W \end{array} \right)=
\frac{1}{\sqrt{g^2+g'^2}}\left(\begin{array}{cc}g&-g'\\ g' & g\end{array}\right)
\left(\begin{array}{c} B\\W \end{array} \right),
\end{equation}
with the fermionic interaction Lagrangian for the gauge fields devolving to
\begin{eqnarray}
{\cal L} &=& -\overline{\psi_L}\gamma.(gW+g'B)\psi_L/2-\overline{\psi_R}g'\gamma.B\psi_R
\nonumber \\
&=&-e\bar{\psi}\gamma.A\psi-e\cot 2\theta\, \overline{\psi_L}\gamma.Z\psi_L
+e\tan\theta\,\overline{\psi_R}\gamma.Z\psi_R.
\end{eqnarray}
It is worth noting that, because $\theta \simeq 30^o$ experimentally to a rather good
approximation, the interaction (at {\em low} energy) almost simplifies to 
\begin{eqnarray}
{\cal L}&\simeq& -e\bar{\psi}\gamma.A\psi+e(\overline{\psi_R}\gamma.Z\psi_R
  - \overline{\psi_L}\gamma.Z\psi_L)/\sqrt{3} 
  = -e\bar{\psi}\gamma.(A-i\gamma_5 Z/\sqrt{3})\psi \nonumber \\
  &\simeq& -(\overline{\psi_L},\overline{\psi_R})\gamma. \left(\begin{array}{cc}
     eA+e'Z & 0\\0& eA-e'Z\end{array}\right)
     \left( \begin{array}{c}\psi_L\\\psi_R\end{array}\right); \quad e' \simeq e/\sqrt{3}.
  \end{eqnarray}
  From this perspective we can view $eA$ and $e'Z$ as forming a part of a U(2) group
  within a matrix $eA+e'\underline{Z}.\underline{\sigma}$ of which only the $Z_3$
  component is in play. However it needs emphasizing that at the level of (7) the 
  left and right gauge fields are {\em not} rotations amongst one another.
  [As we will presently see, we are forced to suppose that $e$ and
  $e'$ coincide at high energy in order to respect universal gravitational coupling to matter.]
It only remains to `curve' the results (4) and (6) by including the vierbein $e_a^m(x)$ and 
metric $g^{mn}(x)$ into the contractions over indices and we magically get the 
interactions of the fermion and boson sources with gravity as well.

The job now is to try to reproduce these conventional constructs through an extended
spacetime-property metric.

\section{\label{sec3}Chiral property}

The very different behaviours of the chiral fermion components under the internal 
gauge transformations point to the need for distinct property coordinates; so at the 
very least we require two independent attributes. In an earlier paper\cite{SD} we treated 
the case of two properties but we assumed that they referred to `up' and `down' 
parts of an isotopic doublet, which led to the parity-conserving unification of 
Yang-Mills U(2) with gravity; there we encountered a property curvature invariant 
$\bar{\zeta}\zeta\equiv\bar{\zeta}_1\zeta^1+\bar{\zeta}_2\zeta^2$. Now it is no longer
necessary to insist upon full U(2) invariance and/or parity conservation, for when we 
identify $\zeta^1\equiv \zeta_L$ with left-handedness and $\zeta^2\equiv \zeta_R$ 
with right handedness, we can permit separate U(1)$_L\times$U(1)$_R$ invariants 
$\overline{\zeta_L}\zeta_L$ and $\overline{\zeta_R}\zeta_R$. Our first task is to 
check that the flat space results can be reproduced, namely a kinetic term and 
{\em no} mass term, after we integrate over attribute space; the next task will be 
to include the scalar fields to recover induced masses and our third task will be 
to obtain the gravitational and gauge field interactions via an extended 
spacetime-property vielbein which `curves' the space. At the end of the day 
we must recognize that the unification of parity violating interactions with gravity will 
not change an ugly duckling (the standard model) into a beautiful swan.

Begin with a Dirac fermion superfield $\Psi(\zeta,\bar{\zeta})$ , expanded in the two 
chiral properties ($\zeta_L$ and $\zeta_R$) which has been made 
anti-selfdual\cite{RD} :-
\begin{eqnarray}
2\Psi&=&(\overline{\zeta_L}\psi_L +{\psi^c}_L\zeta_L)(1-\overline{\zeta_R}\zeta_R)
+ (\overline{\zeta_R}\psi_R +{\psi^c}_R\zeta_R)(1-\overline{\zeta_L}\zeta_L)
\nonumber\\
&=&(\overline{\zeta_L}\psi_L +{\psi^c}_L\zeta_L
+ \overline{\zeta_R}\psi_R +{\psi^c}_R\zeta_R)(1-\bar{\zeta}\zeta).
\end{eqnarray}
As explained in an earlier paper, the adjoint superfield must be defined as
\begin{equation}
2\bar{\Psi} = (\overline{\zeta_L}\overline{{\psi^c}_L}-\overline{\psi_L}\zeta_L
 +\overline{\zeta_R}\overline{{\psi^c}_R}-\overline{\psi_R}\zeta_R)
  (1-\bar{\zeta}\zeta).
\end{equation}
Because of spinorial orthogonality this means that a mass-like term,
\[
2 \bar{\Psi}\Psi = \overline{\zeta_R}\zeta_L\overline{\psi_L}\psi_R +(L\leftrightarrow R),
\] 
will automatically give zero when integrated over property space as there are
insufficient numbers of $\zeta$s. However a
kinetic term, 
\[
2 \bar{\Psi}i\gamma.\partial\Psi = \overline{\zeta_L}\zeta_L(1-2\overline{\zeta_R}\zeta_R)
 \,\overline{\psi_L}i\gamma.\partial\psi_L + (L\leftrightarrow R),
\]
is perfectly fine as it survives integration over property. It is very encouraging that we 
require an extra scalar superfield to induce a mass interaction since it reverberates 
with gauge invariance requirements. 

Thus consider the real anti-selfdual superscalar :
\begin{equation}
\Phi(\zeta,\bar{\zeta})= \phi(1-\overline{\zeta_R}\zeta_R\overline{\zeta_L}\zeta_L)
+ h\overline{\zeta_L}\zeta_R + h^\dag\overline{\zeta_R}\zeta_L 
 + \varphi(\overline{\zeta_L}\zeta_L-\overline{\zeta_R}\zeta_R).
\end{equation}
When interacting with $\Psi$, this will include a piece
\begin{equation}
\bar{\Psi}\Phi\Psi \supset  \overline{\zeta_R}\zeta_L\overline{\psi_L}\psi_R 
\,h\overline{\zeta_L}\zeta_R +{ \rm~h.c.}
\end{equation}
A vacuum expectation value for the chargeless field $\langle h\rangle$ does
duty for a mass term; it will also make an appropriate mixture of gauge bosons 
massive when we introduce curvature into property space.

\section{\label{sec5}Gauge and Gravity extended metric}

The most important objective is to get the correct coupling of gauge fields and
gravity to the chiral fermions through a generalized spacetime-property metric.
The rest (coupling to bosons, gauge-gravity Lagrangian, etc.) will take care of
itself. In a previous paper\cite{SD}, given two properties,  we proved that with a 
metric field $G_{MN}$ in full $x-\zeta$ space having components
\begin{eqnarray*}
G_{mn}&=&g_{mn}[1+c_1\bar{\zeta}\zeta+c_2(\bar{\zeta}\zeta)^2]
   + e^2 l^2\bar{\zeta}(W_mW_n+W_nW_m)\zeta(1+c_3\bar{\zeta}\zeta)/2,\\
G_{m\nu}&=&iel^2(\bar{\zeta}W_m)^{\bar{\nu}}(1+c_3\bar{\zeta}\zeta)/2, \quad{\rm etc.}
\end{eqnarray*}
we could reproduce all gravitational plus gauge field interactions reliably via
the total superscalar curvature $\cal R$, possessing full U(2) invariance. Such
a metric resulted from frame vectors with components
\begin{equation}
{{\cal E}_m}^\alpha = -ie(1+c_3\bar{\zeta}\zeta/2)(W_m\zeta)^\alpha,\quad
{{\cal E}_m}^{\bar{\alpha}} = ie(1+c_3\bar{\zeta}\zeta/2)(\bar{\zeta}W_m)^{\bar{\alpha}},
{\rm ~~etc.}
\end{equation}

In our case we do not require full U(2) invariance but merely U(1)$\times$U(1) 
symmetry, as dictated by experiment for charged leptons; so we can relax the 
condition that the curvature coefficients $c$ and the couplings be the same for the two 
associated gauge fields $A$ and $Z$.  We have a lot more freedom in constructing
the curvatures because $\bar{\zeta}_L\zeta_L$ and $\bar{\zeta}_R\zeta_R$ are
separately invariant. [The appendix contains a full discussion of all property curvatures 
that do not disturb the U(1)$\times$U(1) symmetry and the complications that ensue.] 
In particular the vielbeins ${E_a}^{\zeta_L}, {E_a}^{\zeta_R}$ given there
have forms which ensure that the couplings of the gauge fields
to the fermions correctly sum to $e_L\bar{\psi}_L\gamma.W_L\psi_L+
e_R\bar{\psi}_R\gamma.W_R\psi_R$. Then reinterpreting $e_LW_L=eA+e'Z$,
$e_RW_R=eA-e'Z$, we can indeed reproduce (7). This is all well and good, but we
shall presently see that universal gravitational coupling to the stress tensors of the
gauge fields imposes a strong constraint on the curvature coefficients.

The appendix contains a full treatment of all possible property curvatures, 
consistent with the gauge symmetry. Identifying $c_4$ with $c_L$ and $c_5$ 
with $c_R$,  the metric $G_{m\nu}$ in the $x-\zeta$ sector equals
\[
-i[l^2\bar{\zeta}_Le_LW_L(1+c_L\bar{\zeta}\zeta)
+\ell^2e_RW_R(1+c_R\bar{\zeta}\zeta)] /2.
\]
Under the two independent local coordinate transformations,
\[
\zeta_L\rightarrow\zeta_L'= e^{i\theta_L}\zeta_L,\quad
\zeta_R\rightarrow\zeta_R'= e^{i\theta_R}\zeta_R,
\]
it is readily checked, from the rules of transformation of the metric tensor,
that these simply correspond to independent gauge transformations,
\[
e_LW_L'=e_LW_L+\partial\theta_L,\quad
e_RW_R'=e_RW_R+\partial\theta_R.
 \]
 In turn this translates into
 \[
 eA\rightarrow eA+\partial\theta,\quad e'Z\rightarrow e'Z+\partial\theta',
 \]
which are the independent gauge variations of the more standard gauge 
fields $A$ and $Z$. However we are faced with a serious problem when 
we introduce gravitation...

The dilemma which confronts us is best seen by working out the integral over
the superscalar curvature $\cal R$, as summarized in the appendix:
\begin{equation}
{\cal L}\!=\!\!\int \!\!d^2\zeta d^2\bar{\zeta}\sqrt{-G..}{\cal R}\!=\!\sqrt{-g..}
   \left[\frac{4BR^{[g]}}{l^2\ell^2}
   - \frac{e_L^2c_8}{\ell^2}F^{mn}_LF_{Lmn} -\frac{e_R^2c_6}{l^2}F^{mn}_RF_{Rmn}
   + C\right]
\end{equation}
where $C$ denotes a cosmological term, $F_{mn}\equiv W_{n,m}-W_{m,n}$
and $R^{[g]}$ is the purely gravitational scalar curvature. If we then make the 
replacements 
$$e_LW_L=eA+e'Z, \quad e_RW_R=eA-e'Z,$$ 
we need to ensure cancellation of the cross $AZ$ term which, in turn, forces
$c_8/\ell^2 = c_6/l^2$. Finally the gauge field Lagrangians have to 
couple universally to gravity through their stress tensors so we must further set 
$e'=e$ as a proper normalization condition. This is a very strong consequence 
of attaching force coupling constants to gauge fields in the metric itself.

\section{Concluding Remarks}
{\em This need for coupling constant equality seems to be an unwelcome result}. 
But a moment's reflection leads one to the conclusion that it is an inescapable 
feature which arises from merging the gauge fields and gravity in some kind of enlarged 
metric, {\em not just in our scheme but in any other scheme based on the same 
concepts}. The significant point is that at a semiclassical level\footnote{By 
semiclassical we mean that the metric contains no operators like $Q$ or $Y$ 
but only their eigenvalues such as $e$ or $e'$} the coupling constants must 
always accompany the gauge fields, via the frame vectors that latch on to the 
fermions, and this baggage is inevitably carried into the stress tensors that emerge 
from $G$. 

So the question arises: Is this an impasse or not? We would suggest, like many others, 
that the couplings may very well be equal (or are unified) at some high energy scale 
$l\simeq 10^{13}$ GeV or more, associated with $l$, signifying $e^2(l)=e'^2(l)$, but 
that the observed values at low energy $e'\simeq e/\sqrt{3}$ are due to running 
{\em down} in energy to laboratory scales (corresponding to different gauge groups 
with their particle contents). We would go so far as to assert that any theory which 
attempts to unify gravity with the other forces faces the same dilemma and the 
only reasonable solution is to demand equality of all coupling constants at Planckian 
type scales. The corollary is that if the continuation of the constants to low energy 
via the renormalization group does not match experiment, either the particle content 
associated with the gauge group is wrong or, more drastically, the extended metric 
theory is doomed, unless it is regarded as fully quantum and contains the chargelike 
operators within the frame vectors\footnote{We have never before seen the 
incorporation of differential operators of the coordinates even in a quantized metric field.}.

\appendix
\section{Extended Chiral Minkowski Metric}
We extend spacetime with two property coordinates, $\zeta_L$ and $\zeta_R$. This gives superspace coordinates of $X^M = (x^m, \zeta_L, \zeta_R, \bar\zeta_L, \bar\zeta_R)$.
Our starting point to building the metric is the following metric distance for a flat 4+4 dimensional graded manifold: 
\begin{equation}
ds^2 \!=\! dX^AdX^B \mathcal{I}_{BA}\!=\!dx^a dx^b\eta_{ba}
+ \frac{1}{2} l^2 d\zeta_L d\bar{\zeta_L} 
- \frac{1}{2} l^2 d\bar{\zeta_L}d\zeta_L
 +\frac{1}{2} \ell^2 d\zeta_R d\bar{\zeta_R} 
- \frac{1}{2} \ell^2 d\bar{\zeta_R}d\zeta_R.
\end{equation}
This results in the extended Minkowski metric $\mathcal{I}_{AB}$ taking the form:
\begin{equation}
\mathcal{I}_{AB} = 
\left(
\begin{array}{ccccc}
I_{ab} & I_{a\zeta_L} & I_{a\zeta_R} & I_{a \bar\zeta_L} & I_{a \bar\zeta_R} \\
I_{\zeta_L b} & I_{\zeta_L\zeta_L} & I_{\zeta_L\zeta_R} & I_{\zeta_L \bar\zeta_L} 
& I_{\zeta_L \bar\zeta_R} \\
I_{\zeta_R b} & I_{\zeta_R\zeta_L} & I_{\zeta_R\zeta_R} & I_{\zeta_R \bar\zeta_L} 
& I_{\zeta_R \bar\zeta_R} \\
I_{\bar\zeta_L b} & I_{\bar\zeta_L \zeta_L} & I_{\bar\zeta_L \zeta_R} & 
I_{\bar\zeta_L \bar\zeta_L} & I_{\bar\zeta_L \bar\zeta_R} \\
I_{\bar\zeta_R b} & I_{\bar\zeta_R \zeta_L} & I_{\bar\zeta_R \zeta_R} & 
I_{\bar\zeta_R \bar\zeta_L} & I_{\bar\zeta_R \bar\zeta_R} \\
\end{array}
\right)=
\left(
\begin{array}{ccccc}
\eta_{ab} & 0 & 0 & 0 & 0\\
0 & 0 &0 & \frac{1}{2} l^2  & 0 \\
0& 0 & 0 & 0 & \frac{1}{2} \ell^2\\
0 & -\frac{1}{2} l^2 & 0 & 0 & 0\\
0 & 0 &  -\frac{1}{2} \ell^2  &0& 0\\
\end{array}
\right)
\end{equation}
Notice that we have introduced two separate length scales $l$ and $\ell$ in order
to give ourselves complete freedom and wriggle room.

\section{Frame vectors and Gauge fields}
Consider spacetime dependent U(1)$\times$U(1) phase transformations to the property coordinates $\zeta_L$ and $\zeta_R$ as follows:
\begin{align}
x^m \rightarrow x^m, \;\; \zeta_L \rightarrow e^{i \theta_L(x)} \zeta_L, 
\;\;\bar\zeta_L \rightarrow e^{- i \theta_L(x)} \zeta_L, \;\;
\zeta_R \rightarrow e^{i \theta_R(x)} \zeta_R,  
 \;\; \bar\zeta_R \rightarrow e^{- i \theta_R(x)} \bar\zeta_R.
\end{align}
Now if we want our metric $\mathcal{I}_{AB}$ to be a tensor it has to transform correctly.
With $x$ dependent phase changes in property it does {\em not}.  Therefore,
as in the one coordinate case, we need to fix this by including gauge fields. We
do so by introducing the following upper-triangular frame vector:
\begin{equation}
\mathcal{E}_M{}^{A} =
\left(
\begin{array}{ccccc}
e_m{}^a  & - i L_m \zeta_L&- i R_m \zeta_R & i \bar\zeta_L L_m &  i \bar\zeta_R R_m \\
0 & 1  &  0& 0 & 0\\
0 & 0  &  1& 0 & 0\\
0 & 0  &  0& 1 & 0\\
0 & 0  &  0& 0 & 1\\
\end{array}
\right),
\end{equation}
and its inverse or the vielbein (with ${\cal E_M}^A {E_A}^N = {\delta_M}^N)$:
\begin{equation}
E_A{}^{M} =
\left(
\begin{array}{ccccc}
e_a{}^m  &  i L_a \zeta_L &  i R_a \zeta_R &  - i  \bar\zeta_L L_a &  - i \bar\zeta_R R_a \\
0 & 1  &  0& 0 & 0\\
0 & 0  &  1& 0 & 0\\
0 & 0  &  0& 1 & 0\\
0 & 0  &  0& 0 & 1\\
\end{array}
\right),
\end{equation}
whereupon the metric follows from $G_{MN} = (-1)^{AN} \mathcal{E}_M{}^{A} \mathcal{E}_N{}^B \mathcal{I}_{B A}$; explicitly we get,
\begin{equation}
G_{MN} = 
\left(
\begin{array}{ccccc}
g{}_{m}{}_{n}\!\!+\!\!l^2   L_{m}L_{n}\bar\zeta_L\zeta_L\!\!+\!\! 
\ell^2 R_{m} R_{n}\bar\zeta_R\zeta_R  & -\frac{1}{2} i l^2 \bar\zeta_L L_{m} 
& -\frac{1}{2} i \ell^2 \bar\zeta_R R_{m} & -\frac{1}{2} i l^2 L_{m}\zeta_L 
&-\frac{1}{2} i \ell^2 R_{m}\zeta_R \\ 
 -\frac{1}{2} i l^2 \bar\zeta_L L_{m} & 0 & 0 & l^2 /2 & 0 \\ 
-\frac{1}{2} i \ell^2 \bar\zeta_R R_{m} & 0 & 0 & 0 & \ell^2 /2 \\
 -\frac{1}{2} i l^2 L_{m}\zeta_L & - l^2 /2 & 0 & 0 & 0 \\
-\frac{1}{2} i \ell^2 R_{m}\zeta_R & 0 & - \ell^2 /2 & 0 & 0 \\ 
\end{array}
\right).
\end{equation}
It is readily checked from the tensorial transformation of the metric that the 
changes (B.1) correctly yield the gauge variations,
\begin{equation}
L^\prime{}_{m} = L_m + \theta_{L,m}, \;\;\;\;\; R^\prime{}_{m} = R_m + \theta_{R,m}.
\end{equation}
[Note, there is no need to include gauge couplings $e_R$ and $e_L$ with the 
gauge fields $R$ and $L$ at this stage. In the text $R$ and $L$ have been 
replaced by $e_R W_R$ and $e_L W_L$ .]

\section{Curvature invariants}
Representation (B.4) is not the end of the story. We have the freedom to include  terms
in the metric which are U(1)$\times$U(1) invariant, involving $\bar\zeta_L\zeta_L,
\bar\zeta_R\zeta_R$. But in so doing we have to ensure that {\em all} components 
of $G$ transform correctly; after long and careful analysis we find that the nonzero
metric elements can be finally reduced to
\begin{align*}
G_{mn} \!=& g{}_{m}{}_{n} [1\!+\!c_1 \bar\zeta_L\zeta_L\!+\!c_2 \bar\zeta_R\zeta_R 
\!+\! c_3 \bar \zeta_L \zeta_L \bar\zeta_R \zeta_R]\! +\! 
l^2   L_{m}L_{n}\bar\zeta_L \zeta_L (1\!+\!c_4 \bar\zeta_R \zeta_R )\!+\! 
\ell^2   R_{m} R_{n}\bar\zeta_R \zeta_R (1\!+\! c_5 \bar\zeta_L \zeta_L), \\
G_{m\zeta_L}  =&  - i l^2 \bar\zeta_L L_{m} (1+c_4 \bar\zeta_R \zeta_R)/2,\\
G_{m\zeta_R} =& -i \ell^2 \bar\zeta_R R_{m}(1+c_5 \bar\zeta_L \zeta_L)/2,\\
G_{m\bar\zeta_L}  =&  -i l^2 L_{m}\zeta_L(1+c_4 \bar\zeta_R \zeta_R)/2,\\
G_{m\bar\zeta_R} =& -i \ell^2 R_{m}\zeta_R(1+c_5 \bar\zeta_L\zeta_L)/2, \\
G_{\zeta_L\bar\zeta_L}  =&  l^2(1+ c_{6} \bar\zeta_L \zeta_L + 
c_{4} \bar\zeta_R \zeta_R + c_{7} \bar \zeta_L \zeta_L \bar\zeta_R \zeta_R)/2,\\
G_{\zeta_R \bar\zeta_R}  =& \ell^2 (1+ c_{5} \bar\zeta_L \zeta_L + 
c_{8} \bar\zeta_R \zeta_R + c_{9} \bar \zeta_L \zeta_L \bar\zeta_R \zeta_R)/2.
\end{align*} 
The ensuing inverse metric components read:
\begin{align*}
G^{mn} &= g^{mn} [1 - c_1 \bar\zeta_L \zeta_L - c_2 \bar \zeta_R \zeta_R +
 (2 c_1 c_2 - c_3 )\bar\zeta_L\zeta_L \bar\zeta_R\zeta_R],\\
G^{m \zeta_L} &= i L^m \zeta_L (1 - c_2 \bar\zeta_R\zeta_R),\\
G^{m \zeta_R} &= i R^m \zeta_R (1 - c_1 \bar\zeta_L \zeta_L),\\
G^{m \bar\zeta_L} &= - i L^{m} \bar\zeta_L (1-c_2 \bar\zeta_R \zeta_R),\\
G^{m \bar\zeta_R} &= - i R^m \bar\zeta_R (1-c_1 \bar\zeta_L \zeta_L),\\
G^{\zeta_L\bar\zeta_L} &= 2[1-c_6 \bar\zeta_L\zeta_L - c_4 \bar\zeta_R\zeta_R
 + (2 c_6 c_4 - c_7) \bar\zeta_L \zeta_L \bar\zeta_R \zeta_R]/l^2 - 
 L^m L_m \bar\zeta_L\zeta_L (1- c_2 \bar\zeta_R\zeta_R),\\
G^{\zeta_R\bar\zeta_R} &=\!2 [1-c_5 \bar\zeta_L \zeta_L
-c_8\bar\zeta_R\zeta_R+(2c_5 c_8-c_9)\bar\zeta_L\zeta_L\bar\zeta_R\zeta_R]/\ell^2
 -R^m R_m \bar\zeta_R\zeta_R (1 - c_1 \bar\zeta_L \zeta_L),\\
G^{\zeta_L \zeta_R} &= - L^m R_m \zeta_L\zeta_R,\\
G^{\bar\zeta_L\bar\zeta_R} &= - L^m R_m \bar\zeta_L \bar\zeta_R,\\
G^{\zeta_L \bar\zeta_R} &= L^m R_m \zeta_L\bar\zeta_R,\\
G^{\zeta_R \bar\zeta_L} &=  L^m R_m  \zeta_R \bar\zeta_L.\\
\end{align*}
In the text we have substituted $c_4$ by $c_L$ and $c_5$ by $c_R$;
on the other hand, $c_8,c_6$ enter the normalization of the Lagrangians for the 
left and right gauge fields separately.

\section{Metric superdeterminant}

The Berezinian or superdeterminant requires some more work. Following standard 
procedures the superdeterminant is obtained from the graded supermatrix,
\begin{align*}
\sqrt{G_{..}} &= \frac{4\sqrt{g}}{l^2 \ell^2} \bigg[1+ (2 c_1 - c_5 - c_6) \bar\zeta\zeta + 
(2 c_2 - c_4 - c_8) \bar\zeta_R\zeta_R \\
&+ \left(2 c_1 c_2\!+\!2 c_3\!-\! 2 c_1 c_4\!-\!2c_2 c_5\!+\!c_4 c_5\!-\!2c_2 c_6\!+\!
2c_4c_6\!\!-\!\!c_7\!\!-\!\!2c_1c_8\!\!+\!\!2c_5c_8\!\!+\!\!c_6c_8\!\!-\!\!c_9\right)
\bar\zeta_L\zeta_L\bar\zeta_R\zeta_R\bigg].
\end{align*}
The frame vectors ${\cal E}_M{}^{A}$ which produce the metric have the components:
\begin{align*}
{\cal E}_{m}{}^a &= e_m{}^a [1+\frac{c_1}{2} \bar\zeta_L\zeta_L+\frac{c_2}{2} \bar\zeta_R\zeta_R 
+ (\frac{c_3}{2} - \frac{c_1 c_2}{4}) \bar\zeta_L\zeta_L\bar\zeta_R\zeta_R],\\
{\cal E}_{m}{}^{\zeta_L} &=  - i L_m \zeta_L(1+c_4\bar\zeta_R\zeta_R/2),\\
{\cal E}_{m}{}^{\zeta_R} &=  - i R_m \zeta_R(1+c_5\bar\zeta_L\zeta_L/2),\\
{\cal E}_{m}{}^{\bar\zeta_L} &=   i L_m \bar\zeta_L(1+c_4\bar\zeta_R\zeta_R/2),\\
{\cal E}_{m}{}^{\bar\zeta_R} &=   i R_m \bar\zeta_R(1+c_5\bar\zeta_L\zeta_L/2),\\
{\cal E}_{\zeta_L}{}^{\zeta_L} &= {\cal E}_{\bar\zeta_L}{}^{\bar\zeta_L} = 
1+\frac{c_6}{2} \bar\zeta_L\zeta_L + \frac{c_4}{2} \bar\zeta_R\zeta_R + 
(\frac{c_7}{2} - \frac{c_6 c_4}{4}) \bar\zeta_L\zeta_L\bar\zeta_R\zeta_R,\\
{\cal E}_{\zeta_R}{}^{\zeta_R} &= {\cal E}_{\bar\zeta_R}{}^{\bar\zeta_R} = 
1+\frac{c_5}{2} \bar\zeta_L\zeta_L + \frac{c_8}{2} \bar\zeta_R\zeta_R + 
(\frac{c_9}{2} - \frac{c_5 c_8}{4}) \bar\zeta_L\zeta_L\bar\zeta_R\zeta_R.\\
\end{align*}
Also needed are:
\begin{align*}
({\cal E}_{\zeta_L}{}^{\zeta_L}){}^{-1} &= ({\cal E}_{\bar\zeta_L}{}^{\bar\zeta_L}){}^{-1} =  
1-\frac{c_6}{2} \bar\zeta_L\zeta_L - \frac{c_4}{2} \bar\zeta_R\zeta_R - (\frac{c_7}{2}
 - \frac{3 c_6 c_4}{4}) \bar\zeta_L\zeta_L\bar\zeta_R\zeta_R,\\
({\cal E}_{\zeta_R}{}^{\xi}){}^{-1} &= ({\cal E}_{\bar\zeta_R}{}^{\bar\zeta_R}){}^{-1} = 
1-\frac{c_5}{2} \bar\zeta_L\zeta_L - \frac{c_8}{2} \bar\zeta_R\zeta_R - 
(\frac{c_9}{2} - \frac{3 c_5 c_8}{4}) \bar\zeta_L\zeta_L\bar\zeta_R\zeta_R.\\
\end{align*}
As a useful crosscheck, s$\det{\cal E} = \sqrt{G_{..}}/\sqrt{{\rm s}\det I}$, 
so all is as it should be.

\section{The superscalar curvature}
Now to find the Lagrangian. Using the Palatini form of the Ricci scalar\cite{DS} we get:
\begin{align*}
\mathcal{L} &= \int \sqrt{G_{..}}{\cal R} \;d\zeta_L d\bar\zeta_L d\zeta_R d\bar\zeta_R 
= \frac{4 B}{l^2 \ell ^2} R^{[g]} - \frac{c_6}{l^2} R{}^{m n} R{}_{m n}  - 
\frac{c_8}{\ell ^2} L^{mn} L_{m n} + C,
\end{align*}
where $R_{mn}\equiv R_{n,m} - R_{m,n},\quad  L_{mn} \equiv L_{n,m} - L_{m,n}$, 
$$
B = 
c_7+c_9-c_3+c_1 c_4+c_2 c_5-c_4 c_5+c_2 c_6-2 c_4 c_6+c_1 c_8-2 c_5 c_8-c_6 c_8
$$ 
and
\\
\(
C =
 \frac{16}{l^4 \ell ^4}[-6 \ell ^2 c_1 c_3-6 l^2 c_2 c_3+6 \ell ^2 c_1^2 c_4+4 l^2 c_1 c_2 c_4+4 l^2 c_3 c_4-3 l^2 c_1 c_4^2+4 \ell ^2 c_1 c_2 c_5+6 l^2 c_2^2 c_5+4 \ell ^2 c_3 c_5-8 \ell ^2 c_1 c_4 c_5-8 l^2 c_2 c_4 c_5+3 l^2 c_4^2 c_5-3 \ell ^2 c_2 c_5^2+3 \ell ^2 c_4 c_5^2+6 \ell ^2 c_1 c_2 c_6+3 l^2 c_2^2 c_6+6 \ell ^2 c_3 c_6-18 \ell ^2 c_1 c_4 c_6-8 l^2 c_2 c_4 c_6+\frac{9}{2} l^2 c_4^2 c_6-6 \ell ^2 c_2 c_5 c_6+9 \ell ^2 c_4 c_5 c_6-6 \ell ^2 c_2 c_6^2+12 \ell ^2 c_4 c_6^2+6 \ell ^2 c_1 c_7+4 l^2 c_2 c_7-3 l^2 c_4 c_7-3 \ell ^2 c_5 c_7-6 \ell ^2 c_6 c_7+3 \ell ^2 c_1^2 c_8+6 l^2 c_1 c_2 c_8+6 l^2 c_3 c_8-6 l^2 c_1 c_4 c_8-8 \ell ^2 c_1 c_5 c_8-18 l^2 c_2 c_5 c_8+9 l^2 c_4 c_5 c_8+\frac{9}{2} \ell ^2 c_5^2 c_8-6 \ell ^2 c_1 c_6 c_8-6 l^2 c_2 c_6 c_8+6 l^2 c_4 c_6 c_8+6 \ell ^2 c_5 c_6 c_8+3 \ell ^2 c_6^2 c_8-3 l^2 c_7 c_8-6 l^2 c_1 c_8^2+12 l^2 c_5 c_8^2+3 l^2 c_6 c_8^2+4 \ell ^2 c_1 c_9+6 l^2 c_2 c_9-3 l^2 c_4 c_9-3 \ell ^2 c_5 c_9-3 \ell ^2 c_6 c_9-6 l^2 c_8 c_9]
\).\\

This is all very complicated but it demonstrates how many gauge invariant property 
curvature terms work their way into the system --- 9 in our case. One therefore needs 
some guiding principle for cutting them down. One possible suggestion is to make 
all the curvature pieces {\em fully} U($n$) invariant with $n$ properties, {\em except}
 for those products that involve the gauge fields embedded in the
frame vectors; so in our instance one could think of setting
\[
\ell = l,\quad c_1=c_2,\quad c_4=c_5=c_6=c_8,\quad c_7=c_9,
\]
leaving us with only four constants\footnote{Indeed if we were to set $c_4=c_1, 
c_7=c_3\equiv c_1^2b_3$, we could reduce the whole down to 
$\Lambda c^4/8\pi G_N = 8c_1^2b_3/l^4e^2, 8\pi G_N/c^4=2l^2e^2/c_1(b_3-2)$ 
and there would just be an overall property factor $[1+c_1\bar{\zeta}\zeta+ 
2(c_1\bar{\zeta}\zeta)^2]$ multiplying the property flat elements, including the 
gauge field elements $W\zeta$ in the $x-\zeta$ sector.}, 
one scale and one coupling constant. Although
this simplification really needs some fundamental justification, it can be used to
make the four-property case of SU(2)$_L\times$U(1) more manageable and is even
extensible to the direct product with chromodynamics. The main point is this:  
the property invariant curvatures must be severely constrained before one 
can handle the full standard model; we are uncertain how to do so at this stage.

\end{document}